\documentstyle[11pt,newpasp,twoside,epsf]{article}
\def\edcomment#1{\iffalse\marginpar{\raggedright\sl#1\/}\else\relax\fi}
\reversemarginpar

\begin{document}
\title{Disk Accretion at 10 Myr: Results from the TW Hydrae Association}
 \author{James Muzerolle}
\affil{Steward Observatory, 933 N. Cherry Ave, Tucson, AZ 85721}
 \author{Lynne Hillenbrand}
\affil{Department of Astronomy, California Institute of Technology,
Pasadena, CA 91125}
 \author{Nuria Calvet \& Lee Hartmann}
\affil{Harvard-Smithsonian Center for Astrophysics, 60 Garden St.,
Mail Stop 42, Cambridge, MA 02138}
 \author{C\'esar Brice\~no}
\affil{Centro de Investigaciones de Astronomia, M\'erida, Venezuela}

\begin{abstract}
We present an analysis of the accretion properties of the members
of the TW Hydrae association.  The emission line profile of H$\alpha$
provides explicit evidence of an active accretion flow in 3 out
of the 23 observed members of the association; we use radiative transfer
models to derive approximate accretion rates for these 3 objects.
The resulting accretion
rates for the three 10 Myr-old objects ($< 10^{-9} \; M_{\odot}
\, yr^{-1}$) are well over an order of magnitude lower than typical
values in 1 Myr-old T Tauri stars.  The small fraction
of TW Hydrae association objects still accreting (13\%,
compared with $\sim70\%$ in 1 Myr-old regions like Taurus), along with
the very small accretion rates, points to significant disk evolution
over 10 Myr, providing an important constraint on the timescales
for planet formation.
\end{abstract}

\section{Introduction}

The TW Hydrae association (TWA) has come to be identified as one
of the nearest associations of young stars (de la Reza et al. 1989;
Webb et al. 1999; see also relevant
articles in this volume), and hence has garnered considerable
attention in the past few years with the characterization of
its population.  Moreover, its age of $\sim$ 10 Myr
(Webb et al. 1999) makes it an extremely interesting region with which
to study the processes of planet formation, since initial studies have
shown that 10 Myr should be the approximate timescale for the
dissipation of optically thick circumstellar disks (Strom, Edwards,
\& Skrutskie 1993).

An important question related to the dissipation of disks is
whether gas accretion onto the central star, and hence a reservoir
of gas in the disk, can persist for up to 10 Myr, and whether any
such older disks show signs of evolution from their younger cousins.
The timescale for the dissipation of gas disks is crucial
for the formation of giant planets; core accretion models require
up to 10 Myr or more to form giant planets (e.g.  Bodenheimer,
Hubickyj, \& Lissauer 2000).  The presence of active
accretion in any of the TWA members would thus indicate that indeed
in some cases, optically thick gas disks do survive for up to 10 Myr.
TW Hya itself is already a well-known classical T Tauri star,
with attendant accretion disk (Rucinski \& Krautter 1983).
Extending the initial results presented in Muzerolle et al. (2000),
we investigate whether other members of the association also
exhibit accretion signatures, and use these signatures to examine
evidence of disk evolution.

\section{Observations}

Optical echelle spectra of 11 TWA members were obtained with HIRES
on Keck.  The spectral resolution was $\sim 8$ km s$^{-1}$, with a
wavelength coverage of about 6310 - 8750 \AA.  This dataset is
supplemented with Keck ESI spectra of 4 of the 5 remaining members
listed in Webb et al. (1999), the 3 candidates from Sterzik
et al. (1999), and 5 of the 8 new candidates identified
by Zuckerman et al. (2001).  The ESI spectra are of lower resolution
(37.5 km s$^{-1}$), but have a larger wavelength coverage of
about 3900 - 10750 \AA.  Though the resolution is much lower than
the echelle spectra, it is still sufficient to resolve line profiles
formed in magnetically-mediated accretion flows
(FWHM $\sim 200$ km s$^{-1}$), where the gas is undergoing ballistic
infall onto the star (e.g. Muzerolle, Calvet,
\& Hartmann 2001).  Table 1 lists the objects observed,
along with the H$\alpha$ equivalent widths and FWHM line
widths (multiple values indicate measurements from both ESI and
HIRES, respectively), and whether the [OI] $\lambda$6300 line appears
in emission.

\begin{table}
\caption{TWA Members Observed}
\scriptsize
\medskip
\begin{tabular}{lcccc}
\tableline
Object & EW(H$\alpha$) & FWHM(H$\alpha$) & [OI] $\lambda$6300? & comments \cr
& (\AA) & (km s$^{-1}$) & & \cr
\tableline
TW Hya & -238,-165 & 236,203 & $yes$ & \cr
TWA 2 & -1.8 & 73 & $no$ & \cr
Hen 3$-$600A & -26,-27 & 150,163 & $yes$ & double-lined binary\cr
Hen 3$-$600B & -6.7 & 96 & $no$ & ESI only\cr
HD 98800 & filled & - & $no$ & ESI only\cr
TWA 5A & -9,-7.1 & 140,112 & $no$ & double-lined binary\cr
TWA 6 & -4.2 & 142 & - & double-lined binary\cr
TWA 7 & -5.7 & 64 & - & \cr
TWA 8A & -8.5 & 68 & - & \cr
TWA 8B & -12.5 & 64 & - & \cr
TWA 9A & -3.2 & 87 & $no$ & ESI only\cr
TWA 9B & -5.1 & 72 & $no$ & ESI only\cr
TWA 10 & -7.1,-4.5 & 73,65 & $no$ & \cr
TWA 11A & abs. & - & - & \cr
TWA 11B & -3.8 & 59 & - & \cr
TWA 12 & -5.3 & 78 & $no$ & ESI only\cr
TWA 13A & -1.8 & 82 & $no$ & ESI only\cr
TWA 13B & -4.0 & 84 & $no$ & ESI only\cr
TWA 14 & -12.5 & 205 & $no$ & ESI only\cr
TWA 15A & -9.5 & 85 & $no$ & ESI only\cr
TWA 16 & -3.6 & 79 & $no$ & ESI only\cr
TWA 17 & -2.7 & 137 & $no$ & ESI only, broadened absorption lines\cr
TWA 18 & -3.4 & 102 & $no$ & ESI only\cr
\tableline
\tableline
\end{tabular}
\end{table}

\section{The Accretors of TWA}

Only 3 of the 23 TWA members observed, TW Hya, Hen 3-600A, and
TWA 14, show broad, asymmetric Balmer
emission line profiles indicative of active accretion.  All other
objects exhibit narrow (FWHM $\la 100$ km s$^{-1}$) and symmetric
Balmer line profiles, typical of the chromospheric emission of weak
T Tauri stars (a few of these objects are probable spectroscopic
binaries, where the photospheric absorption lines are broadened or
doubled, in which case the Balmer lines are slightly broader at
FWHM $\sim 100-140$ km s$^{-1}$).  Two of the three objects with
Balmer profiles showing gas infall, TW Hya and Hen 3-600A, also show
[OI] $\lambda$6300 emission, an indicator of mass outflow seen only in
actively accreting young stars; none of the remaining stars
observed with ESI show this emission.  Furthermore, these two objects
exhibit infrared excess emission indicative of circumstellar
disks (Jayawardhana et al. 1999a,b)\footnote{No published infrared
observations yet exist for TWA 14.}.  Finally, TW Hya and TWA 14
exhibit continuum veiling of their photospheric
absorption features, with values of 2.4 and 0.2, respectively,
at 4400 \AA (because Hen 3-600A is a double-lined binary,
reliable veiling determinations cannot be made without knowing
the stellar properties of each component).

\begin{figure}
\plotone{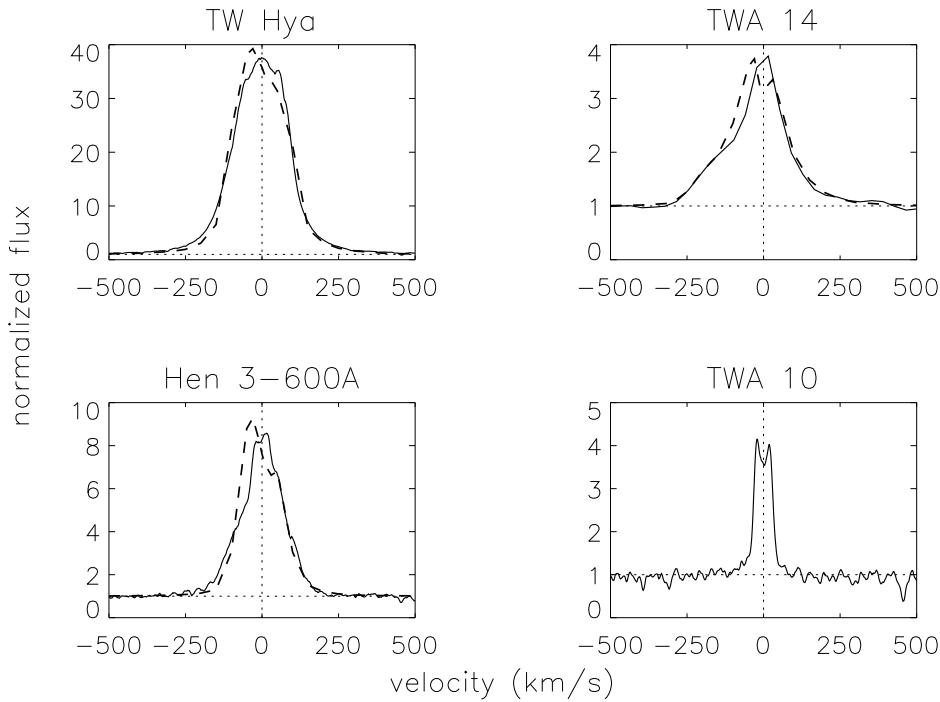}
\caption{Observed (solid lines) and model (dashed lines)
H$\alpha$ profiles for the three TWA accretors, plus the
observed profile of a typical non-accreting object showing
chromospheric emission (TWA 10).}
\end{figure}

In order to quantify the accretion activity, we have modeled
the H$\alpha$ line profiles using radiative transfer models of
magnetospheric accretion.  These models successfully reproduce
the observed line emission in typical 1 Myr-old classical T Tauri
stars in Taurus (Muzerolle, Calvet, \& Hartmann 2001).
Muzerolle et al. (2000) have already presented results for
TW Hya and Hen 3-600A; see that paper for details.  Model
comparisons from that paper are shown in Figure 1, along with
new results for TWA 14.  Note that the stellar parameters
for this object are not well-determined; we assume the same
distance and age as TW Hya, and, given the similar spectral types,
also the same mass and radius.  In each of the three accreting
TWA members, the model accounts very
well for the observed profile shape, width, and strength,
with surprisingly small mass accretion rates ($\dot M = 5\times
10^{-10} \; M_{\odot} \, yr^{-1}$ for TW Hya, and
$\dot M = 5\times 10^{-11}$ for Hen 3-600A and TWA 14)
compared to typical values measured in 1 Myr-old accretors
($\dot M \sim 10^{-8}$).  We note that the two HIRES and ESI
observations of TW Hya and Hen 3-600A show relatively similar
profile shapes and line strengths, with TW Hya showing somewhat
stronger blueshifted absorption in the ESI spectrum.

The H$\alpha$ models can only provide constraints on the mass
accretion rate to within about a factor of 5, however, because
of uncertainties in the parameterized gas temperature
and magnetospheric size (see Muzerolle, Calvet, \& Hartmann 2001).
A much more definitive measurement of $\dot M$
can be made from the luminosity of the UV continuum excess,
which is produced in the accretion shock formed as accreting
material falls onto the surface of the star.  Muzerolle et al. (2000)
used the detailed accretion shock calculations of Calvet \&
Gullbring (1998) to model the UV and optical spectral energy
distribution of TW Hya, the only one of the TWA accretors
with sufficient UV observations.  The resulting value of
the accretion rate, $\sim 4 \times 10^{-10} \; M_{\odot} \,
yr^{-1}$, is in good agreement with the H$\alpha$ models.
\footnote{Batalha and Alencar, this volume, present
extensive observations of the blue excess and line profile
variations, deriving larger accretion rates for TW Hya.
The reason for the discrepancy is unclear, but possibly due
in part to intrinsic variability.}

\section{Accretion at 10 Myr}

Due to the similarity of the Balmer emission line profiles
of the TWA accretors to those in the well-studied younger
objects in Taurus, and the success the accretion models
have in explaining these observations, it is clear that the
standard disk accretion scenario for low-mass young stars
still applies in these older objects.  Thus, some stars
are able to maintain magnetically-mediated accretion flows,
and associated optically thick accretion disks, for at least
10 Myr.  The mass accretion rates in the 10 Myr-old disks
are significantly lower than in typical 1 Myr-old objects, which
indicates that the gas disk surface densities have decreased
substantially over 10 Myr.  Such a decrease is in fact predicted
by models of viscous disk evolution (Hartmann et al. 1998).

Moreover, the fraction of TWA members which are still accreting,
and hence possess optically thick gas disks, is only 13\%.
This is significantly lower than that found in the 1 Myr-old
star forming regions such as Taurus and Orion, where disk
fractions as determined from near-infrared excesses range
from 60-80\% (Kenyon \& Hartmann 1995; Lada et al. 2000).
Both the decreased accretion disk fraction and the very small
mass accretion rates in the three TWA accretors lend substantial
evidence for significant disk evolution occurring over 10 Myr.
Recent results using near- and mid-infrared excesses as
signatures of optically thick inner and outer disk regions
also point to similar disk dissipation timescales
(e.g., Jayawardhana, this volume).

The presence of the three accreting 10 Myr-old T Tauri stars
further begs the question of why their disks have survived for
so long when optically thick gas disks in the remaining TWA members
have since dissipated.
The presence of close companions does
not seem to preclude long-lived accretion disks, since Hen 3-600A
is a spectroscopic binary with accretion signatures, and also
has a third companion (Hen 3-600B, not accreting) $\sim 50$ AU
distant.  As noted by Muzerolle et al. (2000), a circumbinary
disk around the two components of Hen 3-600A should be truncated
to $\sim 20$ AU; for the gas to have lasted for 10 Myr, the
initial mass of the circumbinary disk must have been more massive
than typical ($\ga 0.1 \; M_{\odot}$).  It is possible
that planet formation has led to the dissipation of gas disks
in the other TWA stars, some of which do exhibit mid-infrared
emission from debris disks (Jaywardhana et al. 1999b), which may
represent the detritus of the planet formation process.
Planet formation may be ongoing in the disks of the three accretors;
the spectral energy distribution of TW Hya indicates that dust
coagulation or clearing has occurred within the innermost
$\sim 1$ AU of its disk (see D'Alessio, this volume).

\end{document}